\documentclass{ifacconf}
\usepackage{graphicx}
\usepackage{natbib}
\usepackage[T1]{fontenc}
\usepackage[utf8]{inputenc}

\usepackage{amsmath}
\usepackage{amsfonts}
\usepackage[exponent-product=\cdot]{siunitx}
\DeclareSIUnit{\powerTwo}{\ensuremath{{}^{\circ^2}}}
\DeclareSIUnit{\powerThree}{\ensuremath{{}^{\circ^3}}}
\DeclareSIUnit{\powerFour}{\ensuremath{{}^{\circ^4}}}
\usepackage{xparse}

\usepackage{booktabs}
\usepackage{graphicx}
\usepackage{epstopdf}
\usepackage{cuted}
\usepackage{xcolor}
\definecolor{vgRed}{RGB}{193, 48, 24}
\definecolor{vgOrange}{RGB}{243, 111, 19}
\definecolor{vgYellow}{RGB}{235, 203, 56}
\definecolor{vgGreen}{RGB}{162, 185, 105}
\definecolor{vgLightBlue}{RGB}{13, 149, 188}
\definecolor{vgDarkBlue}{RGB}{6, 56, 81}

\newcommand{\R}{\mathbb{R}}
\newcommand{\Rp}{\R^+}
\newcommand{\Rpz}{\Rp_0}

\usepackage{acronym}
\acrodef{bem}[BEM]{blade element momentum theory}
\acrodef{lq}[LQ]{linear quadratic}
\acrodef{mpc}[MPC]{Model Predictive Controller}
\acrodef{wt}[WT]{Wind Turbine}
\acrodef{pi}[PI]{proportional-integral}
\acrodef{siso}[SISO]{single-input single-output}
\acrodef{lidar}[LIDAR]{light detection and ranging}
\acrodef{del}[DEL]{damage equivalent load}

\usepackage{tikz}
\usepackage{pgfplots}
\usepackage{pgfplotstable}
\pgfplotsset{compat=newest}
\usetikzlibrary{plotmarks}
\usetikzlibrary{fit}
\usetikzlibrary{positioning}
\usetikzlibrary{shapes}
\usetikzlibrary{arrows,automata,plotmarks}
\usetikzlibrary{decorations.pathreplacing,decorations.markings}
\usetikzlibrary{backgrounds}
\usetikzlibrary{calc}
\usetikzlibrary{patterns}
\usetikzlibrary{matrix}
\usepgfplotslibrary{groupplots}
\usepgfplotslibrary{external}
\tikzset{external/system call={pdflatex \tikzexternalcheckshellescape -halt-on-error
    -interaction=batchmode -jobname "\image" "\texsource"}}
\usepackage{todonotes}
\renewcommand{\todo}[2][]{\tikzexternaldisable\@todo[#1]{#2}\tikzexternalenable}

\pgfplotsset{every axis/.append style={semithick,tick style={major tick
            length=4pt,semithick,black}}}

\pgfkeys{/pgfplots/x axis shift down/.style={
        x axis line style={yshift=-#1},
        xtick style={yshift=-#1},
        tick align=outside,
        xticklabel shift={#1}}}

\pgfkeys{/pgfplots/y axis shift left/.style={
        y axis line style={xshift=-#1},
        ytick style={xshift=-#1},
        yticklabel shift={#1},
        scaled y ticks = false,
        tick align=outside,
        y tick label style={/pgf/number format/fixed,
        }
    }
}

\pgfplotsset{myPlot/.style={%
        width=8cm,
        height=4cm,
        line width = 0.7pt,
        separate axis lines,
        axis x line*=bottom,
        x axis shift down = 3pt,
        enlarge x limits=false,
        axis y line*=left,
        y axis shift left = 6pt,
        enlarge y limits={abs=.25pt},
        enlarge x limits={abs=.25pt},
    }
}
\usepackage{mathtools,stackengine,scalerel}
\newcommand{\rawsat}[3]{\ThisStyle{\raisebox{#1\LMpt}{\kern.5\LMpt\scaleto{\rawsatimg{#3}}{#2\LMex}\kern.5\LMpt}}}
\newcommand{\rawsatimg}[1]{%
\tikzexternaldisable
\begin{tikzpicture}
\coordinate (A) at (-7,-7);
\coordinate (B) at (-2,-7);
\coordinate (C) at (2,7);
\coordinate (D) at (7,7);
\draw [black, line width=#1pt] (A)--(B)--(C)--(D);
\end{tikzpicture}%
\tikzexternalenable
}
\makeatletter
\newcommand\sat{
    \relax\if@display
        \mathop{\rawsat{-5}{3.2}{20}}
    \else
        \mathop{\rawsat{-2.4}{2.25}{30}}
    \fi
}

\newcommand{\gettikzxy}[3]{%
  \tikz@scan@one@point\pgfutil@firstofone#1\relax
  \edef#2{\the\pgf@x}%
  \edef#3{\the\pgf@y}%
}

\pgfdeclareshape{satnode}{
\inheritsavedanchors[from={rectangle}]
\inheritbackgroundpath[from={rectangle}]
\inheritanchorborder[from={rectangle}]
\foreach \x in {center,north east,north west,north,south,south east,south west, west, east}{
\inheritanchor[from={rectangle}]{\x}
}
\foregroundpath{
\pgfpointdiff{\northeast}{\southwest}
\pgf@xa=\pgf@x \pgf@ya=\pgf@y
\northeast
\pgfpathmoveto{\pgfpoint{0}{0.45\pgf@ya}}
\pgfpathlineto{\pgfpoint{0}{-0.45\pgf@ya}}
\pgfpathmoveto{\pgfpoint{0.45\pgf@xa}{0}}
\pgfpathlineto{\pgfpoint{-0.45\pgf@xa}{0}}
\pgfpathmoveto{\pgfpointadd{\southwest}{\pgfpoint{-0.2\pgf@xa}{-0.3\pgf@ya}}}
\pgfpathlineto{\pgfpointadd{\southwest}{\pgfpoint{-0.5\pgf@xa}{-0.3\pgf@ya}}}
\pgfpathlineto{\pgfpointadd{\northeast}{\pgfpoint{-0.5\pgf@xa}{-0.3\pgf@ya}}}
\pgfpathlineto{\pgfpointadd{\northeast}{\pgfpoint{-0.4\pgf@xa}{-0.3\pgf@ya}}}
{
   \pgftransformshift{\pgfpointadd{\northeast}{\pgfpoint{-0.4\pgf@xa}{-0.3\pgf@ya}}}
   \pgftransformscale{0.5}
   \pgfsetcolor{black}
}
\pgfusepath{stroke}
}
}

\pgfdeclareshape{lowpassnode}{
\inheritsavedanchors[from={rectangle}]
\inheritbackgroundpath[from={rectangle}]
\inheritanchorborder[from={rectangle}]
\foreach \x in {center,north east,north west,north,south,south east,south west, west, east}{
\inheritanchor[from={rectangle}]{\x}
}
\foregroundpath{
\pgfpointdiff{\northeast}{\southwest}
\pgf@xa=\pgf@x \pgf@ya=\pgf@y
\northeast
\pgfpathmoveto{\pgfpoint{0.4\pgf@xa}{-0.2\pgf@ya}}
\pgfpathlineto{\pgfpoint{0}{-0.2\pgf@ya}}
\pgfpathlineto{\pgfpoint{-0.4\pgf@xa}{0.2\pgf@ya}}
\pgfusepath{stroke}
}
}

\makeatother

\begin{document}
\begin{frontmatter}

\title{LQ Optimal Control for Power Tracking Operation of Wind Turbines}
\thanks{This work was partially supported by the German Federal Ministry for Economic Affairs and and Climate Action (BMWK), project no. 03EE2036C.}

\author[First]{Aaron Grapentin} 
\author[First]{Arnold Sterle} 
\author[First,Second]{Jörg Raisch}
\author[First]{Christian A. Hans}

\address[First]{Control Systems Group, Technische Universität Berlin, Germany (e-mail: \texttt{\{grapentin, sterle, raisch, hans\}@control.tu-berlin.de}).}
\address[Second]{Science of Intelligence, Research Cluster of Excellence, Berlin, Germany}

\begin{abstract}
In this paper, an approach for active power control of individual wind turbines is presented.
State-of-the-art controllers typically employ separate control loops for torque and pitch control.
In contrast, we use a multivariable control approach.
In detail, active power control is achieved by using reference trajectories for generator speed, generator torque, and pitch angle such that a desired power demand is met if weather conditions allow.
Then, a \ac{lq} optimal controller is used for reference tracking.
In an OpenFAST simulation environment, the controller is compared to a state-of-the-art approach.
The simulations show a similar active power tracking performance, while the \ac{lq} optimal controller results in lower mechanical wear.
Moreover, the presented approach exhibits good reference tracking and by improving the reference trajectory generation further performance increases can be expected.
\end{abstract}

%\begin{keyword}
%Wind Turbine, Linear Quadratic Optimal Control, Damage Equivalent Loads
%\end{keyword}

\end{frontmatter}
%===============================================================================
\acresetall

% !TeX spellcheck = en_US
% !TeX encoding = UTF-8
% !TEX root = paper.tex

\section{Introduction}
In the last two decades, renewable energy sources have been scaled up quickly around the world \citep{luz2018multi}.
Apart from installing new units, an effective strategy to increase the share of renewable energy sources is to improve existing renewable generators.
After hydro, wind power counts as the second most popular source of renewable energy \citep{child2019flexible}.
For some markets, e.g., Spain or Denmark, the electricity grid relies heavily on wind power \citep{world2010report}.
In such settings, grid operators require additional services from wind farms to ensure grid stability.
In this context, active power control is an important operation mode, where a wind park does not produce as much power as possible but a given power signal is tracked.
Clearly, active power control requires individual turbines to perform power tracking.
Apart from power tracking performance, aspects, such as mechanical wear on wind turbine components must be considered in the controller design to decrease operation and maintenance costs \citep{Aho2012,kelley2007comparing}.

Traditionally, wind turbine controllers employ separate control loops for the pitch angle and the generator torque \citep{aho2013active}.
Switched nonlinear single-input single-output controllers are typically used to obtain the generator torque based on the generator speed.
For pitch control, a gain-scheduled \ac{pi} controller has been industry standard for many years \citep{jonkman2009definition}.
Adaptions presented in \citet{schlipf2016lidar} led to some improvements but the overall controller structure has largely remained unchanged.

For some time, multivariable control strategies, such as model predictive control or \ac{lq} optimal control, have been researched \citep[see, e.g.,][]{ostergaard2007gain,mirzaei2013}.
The approach presented in \citet{ostergaard2007gain}, focuses on generator modeling and high wind speeds.
The model predictive controller presented in \citet{mirzaei2013} is based on a wind turbine model with five linear states based on uncertain \ac{lidar} measurements.
In this work, we present an \ac{lq} optimal control approach which works throughout all expected wind speeds, while being based on a simple model with only one state.
The model assumes a rigid gearbox with no damping to link rotor and generator.
We show that the \ac{lq} optimal controller obtained from the simplified model performs better than state-of-the-art controllers.
In a numerical case study, the developed controller is benchmarked and evaluated in an OpenFAST simulation environment \citep{jonkman2022openfast} along with a baseline controller.
This includes an analysis of dynamical behavior as well as an investigation of mechanical wear.

The remainder of this paper is organized as follows.
The wind turbine model is provided in Section \ref{sec:model}.
In Section \ref{sec:reference}, a state-of-the-art controller is introduced.
In Section \ref{sec:lqr}, our \ac{lq} optimal control scheme is proposed.
In Section \ref{sec:case_study}, a case study that compares state-of-the-art with our multivariable controller is presented.
Finally, in Section \ref{sec:conclusion}, conclusions and suggestions for future work are provided.

\subsection{Notation}
The set of real numbers is denoted by $\R$ and the set of positive real numbers by $\Rp$.
The set of nonnegative real numbers is denoted by $\Rpz$, while $\mathbb{N}$ denotes the positive integers.
The $\sat$ operator applies a saturation, i.e.,
\begin{align}
y = \sat_{\underline{x}}^{\overline{x}}x
=\begin{cases}
\underline{x}, &\text{if } x \leq \underline{x},\\
\overline{x}, &\text{if } x \geq \overline{x},\\
x, &\text{else},
\end{cases}
\end{align}
where $x\in\R$ is being saturated by the upper bound $\overline{x}\in\R$ and the lower bound $\underline{x}\in\R$, with $\underline{x}<\overline{x}$, such that $y\in[\underline{x}, \overline{x}]$.
The $\operatorname{LUT}$ (lookup table) operator maps a given input using linear interpolation into the codomain based on underlying data points.
Finally, $I_j$, with $j\in\mathbb{N}$, represents the $j$-by-$j$ identity matrix and $0_2 = \begin{bmatrix}0&0\end{bmatrix}^T$.

% !TeX spellcheck = en_US
% !TeX encoding = UTF-8
% !TEX root = paper.tex

\section{Model}
\label{sec:model}
The considered wind turbine model is composed of three parts: rotor aerodynamics, drive train, and generator \citep[see][]{Aho2012,Boersma2017}.
First, the rotor blades convert wind into rotational power.
Then, the drive train converts slow rotations into fast ones.
Finally, the generator converts mechanical into electrical power.
Yaw control is not considered because we assume a constant wind direction.
In what follows, we introduce this model in more detail and derive a linearized version of the overall dynamics.

\subsection{Rotor Aerodynamics}
The wind that passes through the round rotor swept area of a wind turbine with radius $r\in\Rp$ has a power of
\begin{equation}
P_{wind}(t) = \frac{\rho}{2}\pi r^2 V(t)^3,
\label{eq:wind_power}
\end{equation}
where $\rho\in\Rp$ and $V(t)\in\Rp$ denote air density and wind speed at time $t$, respectively.
A wind turbine can only convert a certain proportion of the wind power into rotational power $P_r(t)\in\Rp$.
The power coefficient 
\begin{equation}
C_p\big(\lambda(t), \theta(t)\big) = \frac{P_r(t)}{P_{wind}(t)},
\label{eq:power_coefficient_ratio}
\end{equation}
allows to describe this value.
It is a function of the pitch angle $\theta(t)\in\R$ and the tip-speed ratio
\begin{equation}
\lambda(t)=r\frac{\omega_r(t)}{V(t)},
\label{eq:lambda_definition}
\end{equation}
where $\omega_r(t)\in\Rp$ denotes the angular velocity of the rotor.
Combining (\ref{eq:wind_power}) and (\ref{eq:power_coefficient_ratio}), the power extracted by the rotor can be written as
\begin{equation}
P_r(t) = \frac{\rho}{2}\pi r^2\; V(t)^3C_p(\lambda(t),\theta(t)).
\end{equation}
\begin{figure}[htbp]
	\centering
	\includegraphics{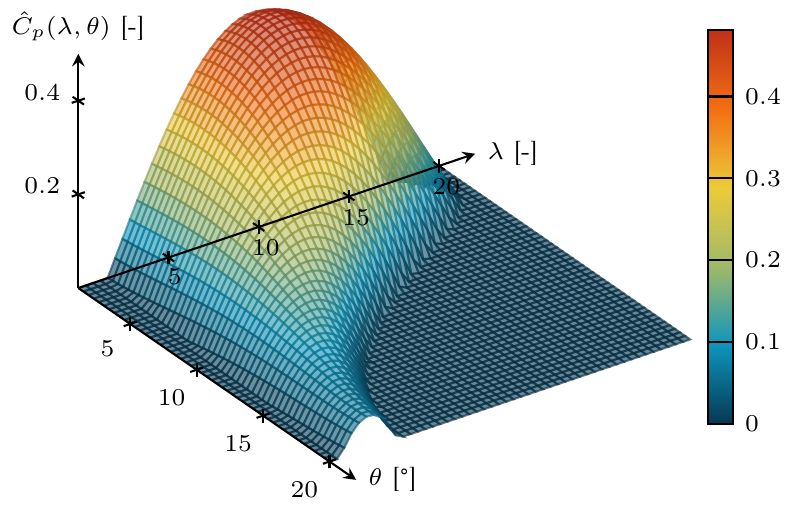}
	\caption{Approximation of power coefficient $\hat{C}_p\left(\lambda(t),\theta(t)\right)$ based on data from \citet{RWT}.}
	\label{fig:power_coefficient_approximation}
\end{figure}
Then, we can obtain the aerodynamic rotor torque as
\begin{equation}
M_r(t)=\frac{P_r(t)}{\omega_r(t)} =\frac{\rho}{2}\frac{\pi r^2}{\omega_r(t)}V(t)^3C_p(\lambda(t), \theta(t)).
\label{eq:generator_torque_definition}
\end{equation}
A power coefficient of $C_p(\cdot)=1$ corresponds to a full conversion of wind power to rotational power.
However, there exists a theoretical limit for the power coefficient, i.e., $C_p(\cdot)<\frac{16}{27}\approx0.59$, which holds for any rotor and blade configuration \citep{Hansen2008}.
The power coefficient depends on various parameters, such as blade shape, material, and weight.
Tools such as OpenFAST \citep{jonkman2022openfast} estimate the aerodynamic torque at the rotor numerically.
In recent literature \citep[see, e.g.,][]{merabet2011torque}, the approximation
\begin{equation}
\tilde{C}_p= \bigg(\frac{a_1}{\lambda_i}+a_2\theta+a_3\bigg)e^{\frac{a_4}{\lambda_i}} \text{ with } \lambda_i = \frac{1}{\frac{1}{\lambda+a_5} + \frac{a_6}{\theta^3+1}},
\end{equation}
is often used.
However, for the wind turbine considered in this paper, an approximation of the form
\begin{align}
	\hat{C}_p(\lambda(t), \theta(t)) &= \max\big(c_1 + c_2\lambda(t) + c_3\theta(t) + c_4\lambda(t)^2\nonumber\\
	&+ c_5\lambda(t)\theta(t) + c_6\theta(t)^2 + c_7\lambda(t)^3\nonumber\\
	&+ c_8\lambda(t)^2\theta(t) + c_9\lambda(t)\theta(t)^2 + c_{10}\theta(t)^3\nonumber\\
	&+ c_{11}\lambda(t)^4 + c_{12}\lambda(t)^3\theta(t) + c_{13}\lambda(t)^2\theta(t)^2\nonumber\\
	&+ c_{14}\lambda(t)\theta(t)^3 + c_{15}\theta(t)^4, 0\big),
	\label{eq:power_coefficient}
\end{align}
was found to be much more accurate.
The approximation $\hat{C}_p(\cdot)$ is fitted from data points $(\lambda, \theta)\in\mathbb{H}\subset\mathbb{R}^2$.
Note that the approach in Section \ref{sec:lqr} works for different approximations of $C_p$ - the shape from (\ref{eq:power_coefficient}) is not required.
In this work, the IEA \SI{3.35}{\mega\watt} turbine is considered, hence the data points provided by \citet{RWT} are used.
A visualization for the approximated power coefficient is depicted in Figure \ref{fig:power_coefficient_approximation}.

\subsection{Drive Train and Generator}
We consider a rigid drive train and gearbox, i.e., $\omega(t) = N_g\omega_r(t)$, where $\omega(t)\in\Rp$ and $N_g\in\Rp$ denote angular generator shaft velocity and gearbox ratio, respectively.
Note that wind turbines without any gearbox, as described in \citet{wagner2020introduction}, are modeled with $N_g=1$.
The dynamics of the drive train are modeled by
\begin{equation}
J_t \dot{\omega}_r(t) =\frac{J_t}{N_g}\dot{\omega}(t) = M_r(t) - N_g\cdot M_g(t),
\label{eq:gearbox_definition}
\end{equation}
where $M_g(t)\in\Rpz$ is the generator torque and ${J_t\in\Rp}$ the moment of inertia.
The combined efficiency of drive train, generator, and power electronics is denoted by ${\eta\in\;]0,1]\subset\Rp}$.
This allows us to calculate the power via
\begin{equation}
P_e(t)=\eta\omega(t)M_g(t).
\label{eq:generator_definition}
\end{equation}

\subsection{State Model}
Combining (\ref{eq:lambda_definition}), (\ref{eq:generator_torque_definition}), (\ref{eq:gearbox_definition}), and (\ref{eq:generator_definition}) results in a nonlinear state model, with state $\omega(t)$, control inputs $\begin{bmatrix}\theta(t) &M_g(t)\end{bmatrix}^T$, and uncertain input $V(t)$, with dynamics
\begin{align}
\dot{\omega}(t) &= f(\omega, \begin{bmatrix}\theta& M_g\end{bmatrix}^T, V)\nonumber\\
&=\frac{\rho\pi r^2 N_g^2}{2 J_t}\frac{V(t)^3}{\omega(t)}\hat{C}_p\Big(\frac{r}{N_g}\frac{\omega(t)}{V(t)}, \theta(t)\Big) - \frac{N_g^2}{J_t}M_g(t).
\label{eq:model}
\end{align}

\subsection{Linearization and Discretization}
Let us assume, that system (\ref{eq:model}) exhibits a unique equilibrium $\omega^s$ for given $(\begin{bmatrix}\theta^s&M_g^s\end{bmatrix}^T, V^s)$ around which the model is now linearized.
At the equilibrium $\dot{\omega}(t)=0$ holds.
At time $t$, $\xi(t) = \omega(t) - \omega^s$ describes the deviation from $\omega^s$, $\mu(t) = \begin{bmatrix}\theta(t)& M_g(t)\end{bmatrix}^T - u^s$ the deviation from $u^s=\begin{bmatrix}\theta^s& M_g^s\end{bmatrix}^T\in\R^2$, and $\nu(t) = V(t) - V^s$ the deviation from $V^s\in\Rp$.
Let us formulate the linearized state model
\begin{equation}
\dot{\xi}(t) = A_\text{c}\xi(t) + B_\text{c}\mu(t) + F_\text{c}\nu(t),
\label{eq:lin_state_space}
\end{equation}
where
\begin{subequations}
\begin{align}
\begin{split}
A_\text{c} &=\left.\frac{\partial f(\omega, \begin{bmatrix}\theta& M_g\end{bmatrix}^T, V)}{\partial\omega}\right|_{\omega^s, \big(\begin{bmatrix}\theta^s& M_g^s\end{bmatrix}^T, V^s\big)},
\end{split}\\
\begin{split}
B_\text{c} &= 
\begin{bmatrix}
\left.\frac{\partial f(\omega, \begin{bmatrix}\theta& M_g\end{bmatrix}^T, V)}{\partial\theta}
\right|_{\omega^s, \big(\begin{bmatrix}\theta^s& M_g^s\end{bmatrix}^T, V^s\big)}\\
\left.\frac{\partial f(\omega, \begin{bmatrix}\theta& M_g\end{bmatrix}^T, V)}{\partial M_g}
\right|_{\omega^s, \big(\begin{bmatrix}\theta^s& M_g^s\end{bmatrix}^T, V^s\big)}
\end{bmatrix}^T,
\end{split}\\
\begin{split}
F_\text{c} &= \left.\frac{\partial f(\omega, \begin{bmatrix}\theta& M_g\end{bmatrix}^T, V)}{\partial V}\right|_{\omega^s, \big(\begin{bmatrix}\theta^s& M_g^s\end{bmatrix}^T, V^s\big)}.
\end{split}
\end{align}
\end{subequations}
For the implementation on digital controllers, a discrete time model is desirable.
We derive this using the forward Euler method with sampling time $T_s\in\Rp$, obtaining
\begin{equation}
\xi(k+1) = A_\text{d}\xi(k) + B_\text{d}\mu(k) + F_\text{d}\nu(k),
\label{eq:discrete_state_space}
\end{equation}
where $A_\text{d} = 1 + T_s A_\text{c}$, $B_\text{d} = T_sB_\text{c}$, and $F_\text{d} = T_sF_\text{c}$.

% !TeX spellcheck = en_US
% !TeX encoding = UTF-8
% !TEX root = paper.tex

\section{Baseline Control Scheme}
\label{sec:reference}
In this section, a state-of-the-art wind turbine controller \citep{RWT,schlipf2016lidar} is presented.
It serves as a baseline for the comparison with the LQ optimal controller presented in Section \ref{sec:lqr}.

\begin{figure}[htbp]
	\centering
	\includegraphics{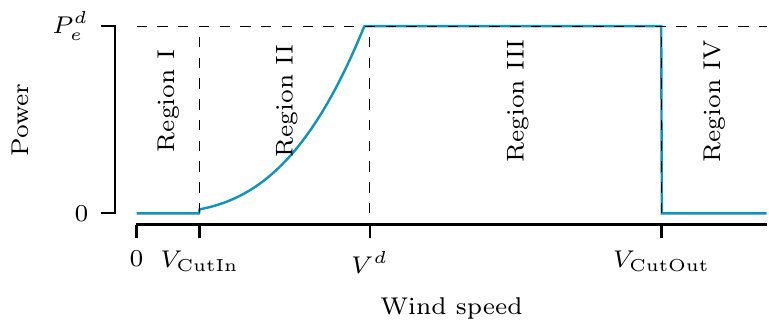}
	\caption{Wind turbine operating regions.}
	\label{fig:regions}
\end{figure}
\begin{figure}[htbp]
	\centering
	\includegraphics{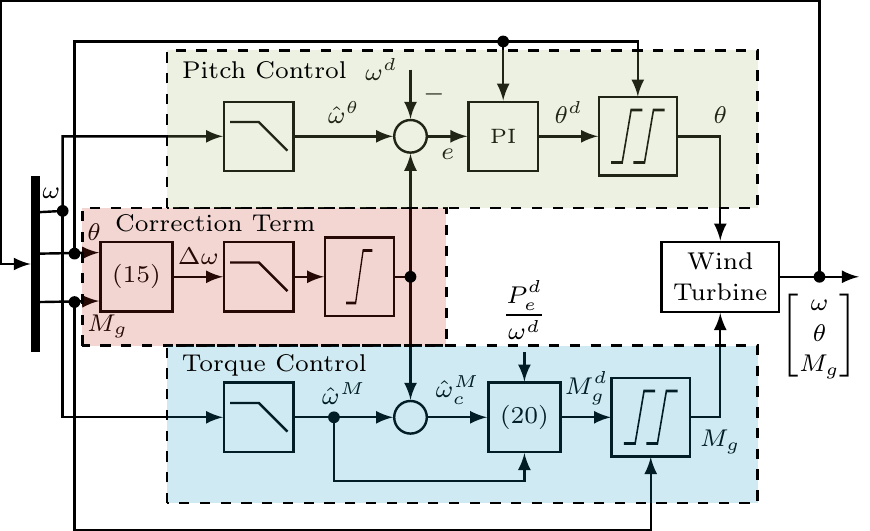}
	\caption{Closed-loop structure of the baseline controller.}
	\label{fig:baseline_diagram}
\end{figure}
Typically, a wind turbine operates in four distinct regions, as displayed in Figure \ref{fig:regions}.
In Regions I and IV, the turbine is turned off due a wind speed below the cut-in wind speed $V_\text{CutIn}$ or above the cut-out wind speed $V_\text{CutOut}$.
Region II is defined by maximization of output power, while the wind speed remains below the wind speed $V^d$ required for the desired power $P_e^d\in\R$.
Finally, in Region III, the wind speed suffices to meet the desired power output $P_e^d$.

The presented baseline controller is capable of the power tracking operation mode, where the turbine provides a desired power below the rated power if weather conditions allow.
Clearly, this is not always possible, since slow wind speeds may not provide enough power to satisfy $P_e^d$.
Power tracking is achieved by computing a desired generator speed $\omega^d(k)$ based on the desired power $P^d_e(k)$, i.e.,
\begin{equation}
	\omega^d(k) = \min\Bigg(\sqrt[3]{\frac{P^d_e(k)}{\eta c_M^*}},\;\omega^\text{Rated}\Bigg).
	\label{eq:correction_term}
\end{equation}
Here, $c_M^*\in\Rp$ is a design parameter and $\omega^\text{Rated}\in\Rp$ is the rated generator speed.
The first term inside the $\min$ operator can be derived from the torque controller, i.e., the last case of (\ref{eq:generator_torque_cases}).
In Figure~\ref{fig:baseline_diagram}, an overview of the baseline controller is shown.
It can be separated into three parts: The general correction term is presented in Section~\ref{sec:baseline_correction}.
The torque control is introduced in Section \ref{sec:ref_torque_control} and the pitch controller is described in Section~\ref{sec:ref_pitch_control}.

\subsection{Correction Term}
\label{sec:baseline_correction}
To smooth out the transition from Region II to Region III, the correction term
\begin{equation}
\Delta\omega(k) = c_\theta^\text{GB}\big(\theta(k-1)-\underline{\theta}\big) + c_M^\text{GB}\big(M_g(k-1)-M_g^\text{Rated}\big),
\end{equation}
is used by the controller.
Here, $c_\theta^\text{GB}, c_M^\text{GB}\in\Rp$ are design parameters while $\underline{\theta},M_g^\text{Rated}\in\R$ denote lower pitch angle bound and rated generator torque, respectively.
Before applying the correction term, a lowpass filter is used, e.g.,
\begin{equation}
	\Delta\hat\omega(k) = (1-\alpha^\text{GB})\cdot\Delta\hat\omega(k-1) + \alpha^\text{GB}\cdot\Delta\omega(k),
\end{equation}
where $\alpha^\text{GB}=\frac{T_s}{T_\text{GB}-T_s}\in[0,1]\subset\Rpz$ is based on $T_s$ and the filter time constant $T_\text{GB}>2T_s$.
The filter is initialized to $\Delta\hat{\omega}(0) = \Delta\omega(0)$.
In what follows, we will describe how the baseline controller is designed based on $\omega^d(k)$ and $\Delta\hat{\omega}(k)$.

\subsection{Torque Control}
\label{sec:ref_torque_control}
For each operating region, the generator torque $M_g(k)$ is computed differently.
The desired generator torque $M_g^d(k)$ is computed based on a lowpass filtered generator speed measurement and the correction term.
The lowpass filtered generator speed $\hat\omega^M(k)$ is derived via
\begin{equation}
\hat\omega^M(k) = (1-\alpha^M)\cdot\hat\omega^M(k-1) + \alpha^M\cdot\omega(k),
\end{equation}
where $\alpha^M=\frac{T_s}{T_M-T_s}\in[0,1]\subset\Rpz$ is based on $T_s$ and the filter time constant $T_M$.
The filter is initialized to $\hat\omega^M(0)=\omega(0)$.
Then, the corrected generator speed
\begin{align}
	\hat{\omega}_c^M(k) = \hat{\omega}^M(k)+\sat_0^\infty\Delta\hat{\omega}(k)
\end{align}
is computed and the desired generator torque is set to be
\begin{equation}
	M_g^d(k) = \begin{cases}
		P_e^d(k)/\omega^d(k),
        &\text{if }\hat{\omega}^M(k) \geq \omega^d(k),\\
		0,
		&\text{else if }\hat{\omega}^M(k) \leq \omega_\text{ci},\\
		c_{12}(\hat\omega^M(k) - \omega_\text{ci}),\!\!\!\!\!\!
        &\text{else if }\hat{\omega}_c^M(k) < \omega_\text{r2},\\
		c_M^*\hat{\omega}^M(k)^2,
		&\text{else if }\hat{\omega}^M(k) < \omega^d(k),
	\end{cases}
	\label{eq:generator_torque_cases}
\end{equation}
where $c_{12}, \omega_\text{ci}, \omega_\text{r2}\in\Rp$ are design parameters.
Before applying the desired generator torque $M_g^d(k)$, a saturation ensures that it lies within bounds $[\underline{M}_g, \overline{M}_g]\subset\R$.
Then, the rate of change is computed via numerical differentiation and limited to the interval $[\underline{\Delta M}_g, \overline{\Delta M}_g]\subset\R$.
By numerical integration, we obtain the generator torque
\begin{equation}
M_g(k) = M_g(k-1)+\sat_{\underline{\Delta M}_g}^{\overline{\Delta M}_g} \Bigg(\sat_{\underline{M}_g}^{\overline{M}_g} \Big(M_g^d(k)\Big) - M_g(k-1)\Bigg).
\end{equation}

\subsection{Pitch Control}
\label{sec:ref_pitch_control}
The desired pitch angle $\theta^d(k)$ is computed based on the lowpass filtered generator speed $\hat{\omega}^\theta(k)$ and the correction term $\Delta\omega(k)$.
The filtered signal is derived through 
\begin{equation}
\hat\omega^\theta(k) = (1 - \alpha^\theta)\cdot\hat\omega^\theta(k) + \alpha^\theta\cdot\omega(k-1),
\end{equation}
where $\alpha^\theta=\frac{T_s}{T_\theta - T_s}\in[0,1]\subset\Rpz$ is based on $T_s$ and the filter constant $T_\theta$.
The filter is initialized to ${\hat\omega^\theta(0)=\omega(0)}$.
The corrected generator speed is then computed as
\begin{align}
	\hat{\omega}_c^\theta(k) =
		\hat{\omega}^\theta(k) + \sat_0^\infty\Delta\hat{\omega}(k)
\end{align}
The desired pitch angle is computed using a PI control loop based on the error
\begin{equation}
e(k) = \hat{\omega}_c^\theta(k) - \omega^d(k).
\end{equation}
The error is integrated using the forward Euler method and saturation limits to obtain the integrated error
\begin{equation}
E(k) = \sat_{\underline{E}(k)}^{\overline{E}(k)}\Big(E(k-1)+T_s e(k)\Big),
\end{equation}
where the lower and upper bound are adapted for each sampling instance $k$, i.e.,
\begin{align}
	&\underline{E}(k) = \frac{\underline{\theta}}{G_K(k)\cdot K_I},
	&\overline{E}(k) = \frac{\overline{\theta}}{G_K(k)\cdot K_I}.
\end{align}
With $G_K(k)=\frac{1}{1+\theta(k-1)/K_K}$ where $K_K\in\Rp$ denotes a design parameter, the integrated error is initialized to ${E(0)=\theta(0)/\big(K_I\cdot G_K(0)\big)}$, where $K_I\in\Rp$ denotes the integral gain.
The proportional gain is denoted by $K_P\in\Rp$.
The desired pitch angle is then obtained as
\begin{equation}
\theta^d(k) = G_K(k)\big(K_P e(k) + K_I E(k)\big).
\end{equation}
Analogously to the torque control, the desired pitch angle and the rate of change are limited by $[\underline{\theta},\overline{\theta}]\in\R$ and $[\underline{\Delta\theta}, \overline{\Delta\theta}]\subset\R$, respectively.
By numerical integration, the pitch angle $\theta(k)$ is then obtained as
\begin{equation}
\theta(k) = \theta(k-1)+\sat_{\underline{\Delta\theta}}^{\overline{\Delta\theta}} \Bigg(\sat_{\underline{\theta}}^{\overline{\theta}} \Big(\theta^d(k)\Big) - \theta(k-1)\Bigg).
\end{equation}

% !TeX spellcheck = en_US
% !TeX encoding = UTF-8
% !TEX root = paper.tex

\section{LQ Optimal Controller}
\label{sec:lqr}
\begin{figure}[htbp]
	\centering
	\includegraphics{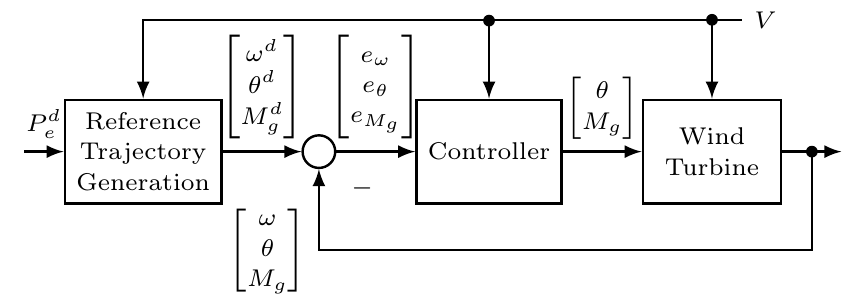}
	\caption{Closed-loop structure of the new control scheme.}
	\label{fig:lq_diagram}
\end{figure}

In this section, a multivariable wind turbine controller is developed.
In Figure~\ref{fig:lq_diagram}, our novel control scheme is shown.
In Section \ref{sec:lqr_augmentation}, we will first augment the model from Section \ref{sec:model}.
In Sections \ref{sec:subsection_lq_control} and \ref{sec:lqr_switching}, a switching \ac{lq} optimal controller is proposed, which enables an operation at all wind speeds.
In Section \ref{sec:lqr_reference}, the reference trajectory generation is presented and in Section \ref{sec:lqr_control_law}, the overall control law is defined.

\subsection{Augmented State Model}
\label{sec:lqr_augmentation}
\begin{figure}[b]
	\centering
	\includegraphics{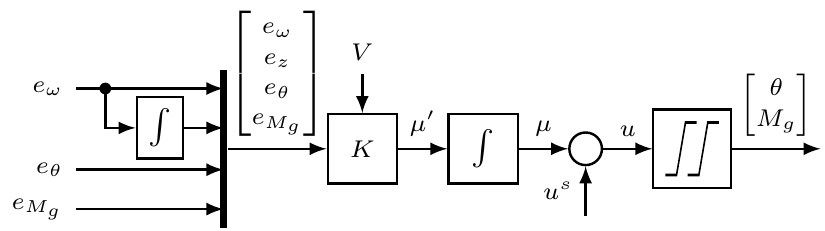}
	\caption{Internal structure of the new controller.}
	\label{fig:control_diagram}
\end{figure}

In Figure~\ref{fig:control_diagram}, a detailed overview of the controller block in Figure~\ref{fig:lq_diagram} is shown.
For accurate tracking of the reference generator speed $\omega^d$, the integral $z(k)$ of the error
\begin{equation}
e_\omega(k) = \omega^d(k) - \omega(k) = \xi^d(k) - \xi(k),
\label{eq:lq_error}
\end{equation}
is added to the model \citep[see, e.g.,][]{birla2015optimal}.
Moreover, the inputs $\mu(k)=\begin{bmatrix}\mu_1(k)&\mu_2(k)\end{bmatrix}^T\in\R^2$ are moved into the state vector and new differential inputs $\mu'(k)\in\R^2$ are added to the model.
This allows to discourage actuation changes that cause mechanical wear in the \ac{lq} optimal control design.
Together with (\ref{eq:lq_error}), this results in the augmented system
\begin{equation}
\begin{bmatrix}\xi(k+1)\\ z(k+1)\\ \mu(k+1)\end{bmatrix}
= \begin{bmatrix}A_\text{d}\xi(k)+B_\text{d}\mu(k) + F_\text{d}\nu(k)\\ z(k) + T_s\xi^d(k) - T_s\xi(k)\\ \mu(k) + T_s\mu'(k)\end{bmatrix}.
\end{equation}
This can be reformulated into
\begin{equation}
x(k+1) = A x(k) +B\mu'(k) +F\tilde{\nu}(k),
\label{eq:lin_ext_state_space}
\end{equation}
with state $x(k)=\begin{bmatrix}\xi(k)& z(k)& \mu(k)^T\end{bmatrix}^T$, external inputs $\tilde{\nu}(k)= \begin{bmatrix}\nu(k)& \xi^d(k)\end{bmatrix}^T$,
and matrices
$$A =
\begin{bmatrix}A_\text{d} & 0 & B_\text{d} \\ -T_s & 1 & 0^T_2 \\ 0_2 & 0_2 & I_2\end{bmatrix},\quad
B =
\begin{bmatrix}0^T_2\\0^T_2\\T_sI_2\end{bmatrix},\quad
F =
\begin{bmatrix}
F_d & 0 \\0 & T_s\\ 0_2 & 0_2
\end{bmatrix}.$$
Note that, the desired generator speed deviation $\xi^d(k)$ is included as an external input, which is provided through the reference trajectory generation block.

\subsection{LQ Optimal Control}
\label{sec:subsection_lq_control}
The \ac{lq} optimal controller is designed for the augmented state model (\ref{eq:lin_ext_state_space}), which is controllable.
The controller uses linear state feedback of the form ${u(k) = -Kx(k)}$, with gain matrix $K\in\mathbb{R}^{p\times n}$, to drive the system to the origin.
To drive the system close to a desired state $x^d(k)\in\R^n$, the control law is modified, into ${u(k) = K(x^d(k) - x(k))}$ \citep{skogestad2005multivariable}.
Matrix $K$ is computed such that the cost function
\begin{equation}
J = \sum_{n=1}^\infty x(k)^TQ\,x(k) + u(k)^TR\,u(k),
\label{eq:cost_function}
\end{equation}
with positive semidefinite $Q\in\R^{4\times 4}$ and positive definite $R\in\R^{2\times 2}$ is minimized.
The values used for $Q$ and $R$ can be found in Tab.~\ref{tab:LQParameters}.
We can find a solution via the discrete-time Riccati equation
\begin{equation}
A^TS\,A-S-A^TS\,B(B^TS\,B+R)^{-1}B^TS\,A+Q=0.
\label{eq:riccati_equation}
\end{equation}
Solving (\ref{eq:riccati_equation}) provides $S\in\R^{4\times 4}$ from which we obtain
\begin{equation}
K=(B^TS\,B+R)^{-1}B^TS\,A.
\end{equation}

\subsection{Switching Controller}
\label{sec:lqr_switching}
Due to distinct operating regions (see Figure \ref{fig:regions}), two gain matrices $K_1$ and $K_2$ are computed based on the linearized model for two different equilibrium points.
Additionally, the cost function is different for both gain matrices, in particular, for $K_1$, the generator speed must be tracked very accurately to maximize power production.
For $K_2$, the cost function is adjusted such that the generator speed and torque are tracked accurately to satisfy the power demand, this is reflected in the weight matrices in Tab.~\ref{tab:LQParameters}.
The gain matrix $K(k)\in\R^{2\times 4}$ is then determined by switching between $K_1$ and $K_2$ based on a hysteresis.
Once the wind speed rises above the upper wind speed threshold $\overline{V}\in\Rp$, $K_2$ is selected as the active gain matrix.
As soon as the wind speed drops below a lower wind speed threshold $\underline{V}\in\Rp$ with $\underline{V}<\overline{V}$, $K_1$ is selected, i.e.,
\begin{align}
K(k) = \begin{cases}
K_1\quad&\text{if } V(k) < \underline{V},\\
K(k-1) &\text{if } \underline{V} \leq V(k) \leq \overline{V},\\
K_2&\text{if } \overline{V} < V(k).\end{cases}
\end{align}
This completes the controller implementation.
Before the controller can be used, reference trajectories need be generated.

\subsection{Reference Trajectories}
\label{sec:lqr_reference}
The desired state
$$x^d(k)=\begin{bmatrix} \omega^d(k) & z^d(k) & \theta^d(k) & M_g^d(k) \end{bmatrix}^T -\begin{bmatrix}\omega^s & 0 & \theta^s & M_g^s\end{bmatrix}^T,$$
contains the reference generator speed $\omega^d(k)$, the reference integral error $z^d(k)$, the reference pitch angle $\theta^d(k)$, the reference generator torque $M_g^d(k)$, and the equilibrium $(\omega^s,\begin{bmatrix}\theta^s& M_g^s\end{bmatrix}^T)$.
Clearly, no integral error is desired, i.e. $z^d(k)=0$.
The derivation of the remaining references will be discussed in what follows.

\subsubsection{Reference Generator Speed:}
\label{sec:lqr_ref_rotorSpeed}
The reference generator speed is the minimum of $\omega^*(k)$ and the power setpoint speed $\omega^{sp}(k)$, i.e.,
\begin{equation}
\hat{\omega}^d(k) = \min\big(\omega^*(k),\, \omega^{sp}(k)\big).
\end{equation}

Here, $\omega^*(k)$ is found by maximizing the power coefficient (\ref{eq:power_coefficient}) for the current pitch angle $\theta(k)$ and wind speed $V(k)$. Thus, the problem
\begin{subequations}
\begin{align}
\max_{\omega(k)}&\quad\hat{C}_p\bigg(\frac{r}{N_g}\frac{\omega(k)}{V(k)}, \theta(k)\bigg),\\
\text{s.t.}&\quad\Big(\frac{r}{N_g}\frac{\omega(k)}{V(k)}, \theta(k)\Big)\in\mathbb{H},
\end{align}
\end{subequations}
is solved to obtain $\omega^*(k)$.
The power setpoint speed $\omega^{sp}(k)$ is computed using a lookup table depending on $P^d_e(k)$, i.e.,
\begin{equation}
\omega^{sp}(k) = \operatorname{LUT}\big(P^d_e(k)\big).
\label{eq:setpoint_lookup_table}
\end{equation}
Analogously to \citet{kim2018design} and \citet{jeon2021design}, steady state simulation data is used to deduce the lookup table.
Here, steady state equilibria tuples of generator speed and output power are aranged in a grid.
This provides an adaptable basis and realizes a simple to use while effective method to convert the desired power into a reference trajectory for the generator speed.

For the reference generator speed $\omega^d(k)$, a lowpass filter is applied to $\hat{\omega}^d(k)$, i.e.,
\begin{equation}
\omega^d(k) = (1-\alpha_1)\cdot\omega^d(k-1) + \alpha_1\cdot\hat{\omega}^d(k),
\end{equation}
where $\alpha_1=\frac{T_s}{T_1+T_s}\in[0,1]\subset\Rpz$ is based on $T_s$ and the filter time constant $T_1$, with $\omega^d(0) = \hat\omega^d(k)$.

\subsubsection{Reference Pitch:}
The reference pitch angle
\begin{equation}
\hat\theta^d(k) = \operatorname{LUT}\big(P^d_e(k), V(k)\big),
\end{equation}
is computed with a lookup table from the same steady state simulation data as in (\ref{eq:setpoint_lookup_table}).
To prevent fast changes $\hat\theta^d(k)$ is lowpass filtered via
\begin{equation}
    \theta^d(k) = (1-\alpha_2)\cdot\theta^d(k) + \alpha_2\cdot\hat\theta^d(k).
\end{equation}
Here, $\alpha_2=\frac{T_s}{T_2+T_s}\in[0,1]$, is based on $T_s$ and the filter time constant $T_2$, with $\theta^d(0) = \hat\theta^d(0)$.

\subsubsection{Reference Generator Torque:}
The reference generator torque is computed such that once $\omega^d(k)$ is reached, exactly the demanded power is produced.
Assuming a perfect generator speed tracking, solving (\ref{eq:generator_definition}) for $M_g(k)$ results in
\begin{equation}
M_g^d(k) = \frac{P_e^d(k)}{\eta\omega^d(k)}.
\label{eq:desired_generator_computation}
\end{equation}

\subsection{Control Law}
\label{sec:lqr_control_law}
Finally, the plant is controlled with the input $u(k)$ which is derived by integrating $\mu'$, i.e.,
\begin{equation}
u(k) = \mu(k) + u^s(k) = \mu(k-1) + T_s\cdot\mu'(k) + u^s(k),
\end{equation}
where $u^s(k)$ denotes the steady state inputs $\begin{bmatrix}\theta^s& M_g^s\end{bmatrix}^T$ of the currently active controller.
The differential input $\mu'(k)$ is computed by the gain matrix $K(k)$ and the error from the reference $x^d(k)$, i.e.,
\begin{equation}
\mu'(k) = K(k)\cdot\big(x^d(k) - x(k)\big).
\end{equation}
Combined into one, the control input $u(k)$ is obtained from
\begin{equation}
u(k) = \mu(k-1) + T_s\cdot K(k)\cdot\big(x^d(k) - x(k)\big) + u^s(k).
\end{equation}
Before applying the inputs to the plant, saturation and slew rate constraints are enforced
\begin{align}
    \theta(k) &= \theta(k-1) + \sat_{\underline{\Delta\theta}}^{\overline{\Delta\theta}}\bigg(\sat_{\underline{\theta}}^{\overline{\theta}}\big(\theta^t(k)\big) - \theta(k-1)\bigg),\\
    M_g(k) &= M_g(k-1) + \\
	&\quad\quad\quad\quad\sat_{\underline{\Delta M}_g}^{\overline{\Delta M}_g}\bigg(\sat_{\underline{M}_g}^{\overline{M}_g}\big(M_g^t(k)\big) - M_g(k-1)\bigg).
\end{align}

% !TeX spellcheck = en_US
% !TeX encoding = UTF-8
% !TEX root = paper.tex

\section{Case Study}
\label{sec:case_study}
In this section, the LQ optimal controller, from Section~\ref{sec:lqr}, is compared with the baseline controller, from Section~\ref{sec:reference}.
The IEA \SI{3.35}{\mega\watt} wind turbine \citep{RWT} is used for all simulations.
The corresponding turbine parameters, such as bounds on the control inputs, their rates of change, and others are presented in Table~\ref{tab:generalParameters}.
In Table~\ref{tab:BaselineParameters}, the baseline controller parameter values from \citet{RWT} are listed.
In Table~\ref{tab:LQParameters}, the LQ optimal controller parameters are listed,
and in Table~\ref{tab:coefficientParameters}, the parameters for (\ref{eq:power_coefficient}) are provided.
These were determined based on the data points provided by \citet{RWT}.
\begin{figure}[b]
	\centering
	\includegraphics{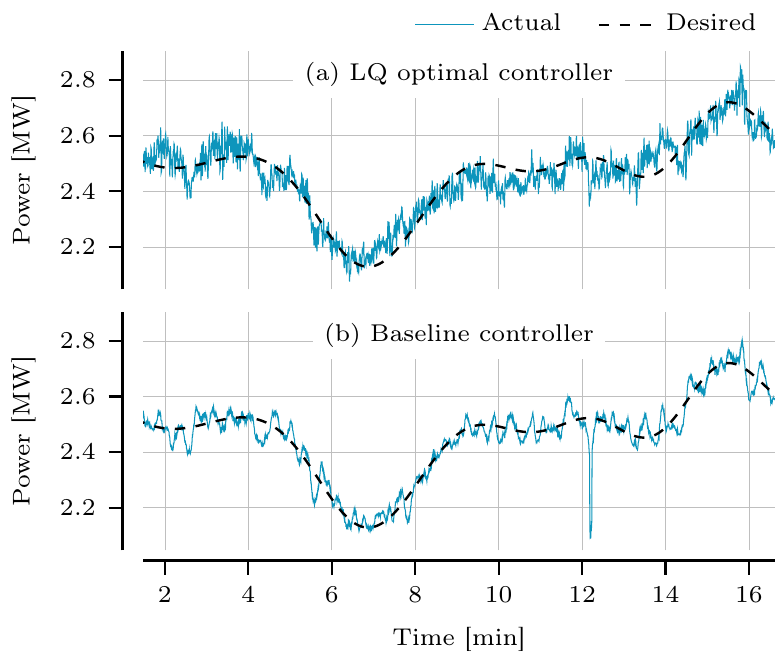}
	\caption{Output power trajectories.}
	\label{fig:powerTracking}
\end{figure}
All simulations are executed with a sampling time of ${T_s=\SI{4}{\milli\second}}$.
From each simulation, the first \SI{1.5}{\minute} are omitted to disregard transient simulation effects.

Power tracking in Region III is evaluated with an average wind speed of \SI[per-mode=fraction]{15}{\meter\per\second} and a turbulence intensity of \SI{9}{\percent}.
In Figure~\ref{fig:powerTracking}, the output power of both controllers, as well as the desired power are shown.
The \ac{lq} optimal controller and the baseline controller both achieve similar root-mean-square (RMS) tracking errors of \SI{45.1}{\kilo\watt} and \SI{44.1}{\kilo\watt}, respectively.
This represents a relative error below \SI{2}{\percent} of the rated power.
A notable difference between the two controllers appears around $t=\SI{12.2}{\minute}$, where the baseline controller dips far below the desired power.
In contrast, the LQ optimal controller remains much closer to the desired value.
\begin{figure}[htbp]
	\centering
	\includegraphics{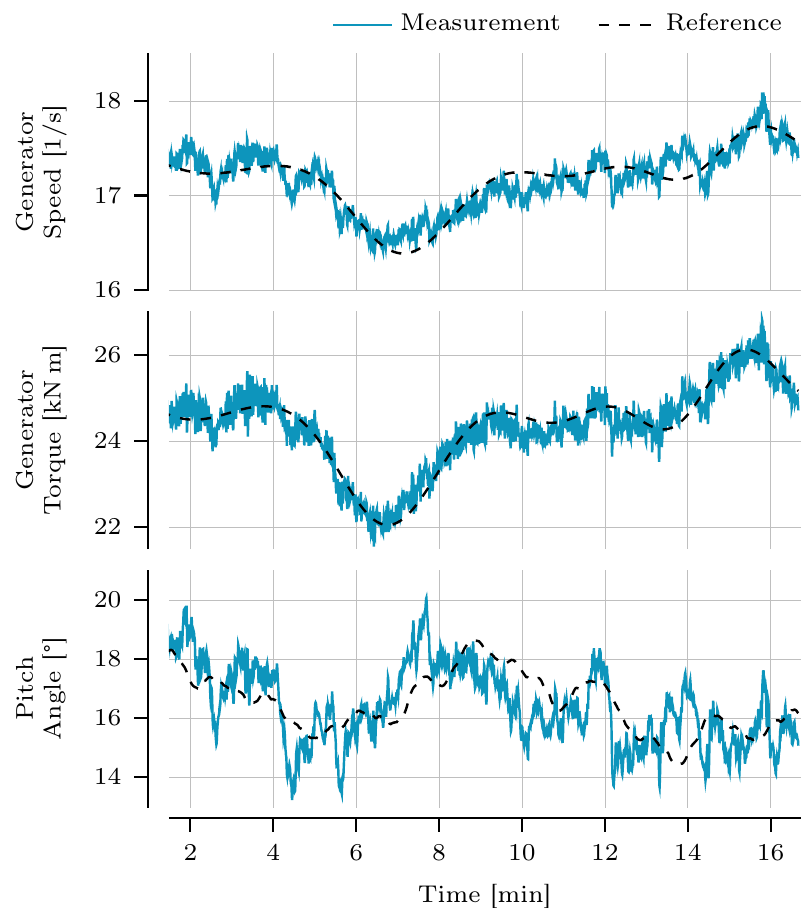}
	\caption{Reference tracking plots at \SI[per-mode=symbol]{15}{\meter\per\second} wind speed with \SI{9}{\percent} turbulence intensity.}
	\label{fig:referenceTrackingR3}
\end{figure}

Figure~\ref{fig:referenceTrackingR3} presents the measurements and reference signals for generator speed, generator torque, and pitch angle for the \ac{lq} optimal controller from the same simulation as in Figure~\ref{fig:powerTracking}.
The three reference trajectories are tracked with different accuracies.
The generator speed and pitch angle references are only tracked approximately.
In contrast, the generator torque signal tracks the reference very well due to its relatively high weight in $Q_2$ (see Table~\ref{tab:LQParameters}).

\begin{figure}[b]
	\centering
	\includegraphics{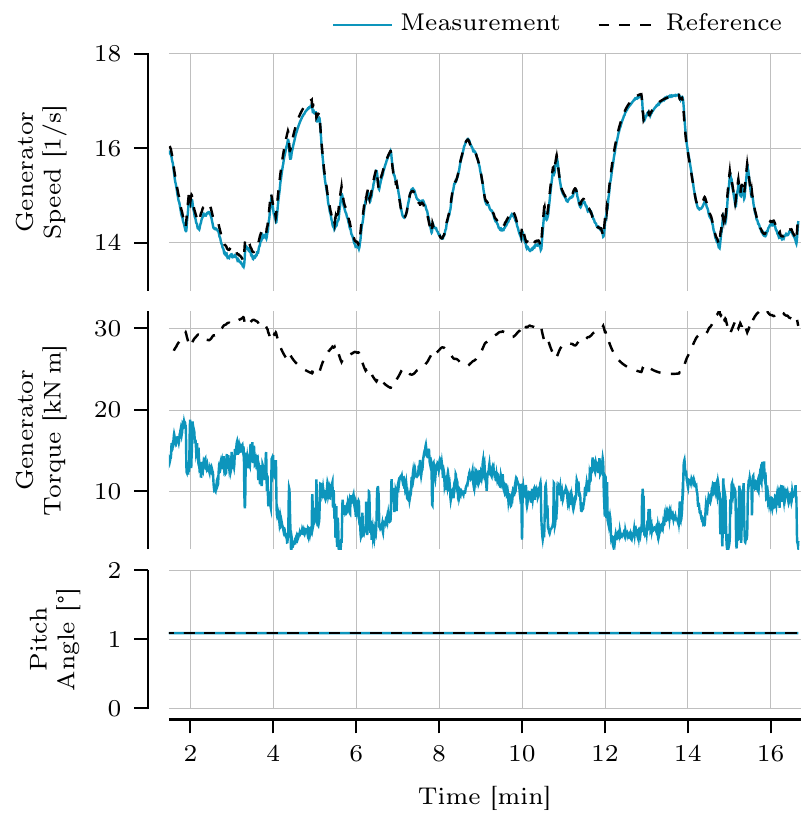}
	\caption{Reference tracking plots at \SI[per-mode=symbol]{6.3}{\meter\per\second} wind speed with \SI{9}{\percent} turbulence intensity.}
	\label{fig:referenceTrackingR2}
\end{figure}
Typically in Region II, the wind speed does not allow for power tracking of arbitrary power demands.
In Figure~\ref{fig:referenceTrackingR2}, measurements and reference signals for generator speed, generator torque, and pitch angle are shown for a wind speed of \SI[per-mode=fraction]{6.3}{\meter\per\second} with a turbulence intensity of \SI{9}{\percent}.

The tracking behavior differs vastly from the results in Figure~\ref{fig:referenceTrackingR3}.
The generator speed and pitch angle are tracked very closely but the generator torque deviates from its reference.
In detail, pitch angle reference and measurement remain at the optimal value of \SI{1.09}{\degree}, which originates from the blade geometry.
Tracking the reference generator speed closely is important to remain at the optimal tip-speed ratio.
In combination with the optimal pitch angle, this allows to extract maximum power from the wind.
Due to low wind speeds, which are not sufficient to achieve the desired output power, the torque of the generator cannot be tracked accurately without decelerating the turbine as thereby deviating from the generator speed reference.
To stop the turbine from slowing down, deviations are acceptable and desired, hence the low weight in the fourth entry in $Q_1$ (see Table~\ref{tab:LQParameters}).

\begin{figure}[b]
	\centering
	\includegraphics{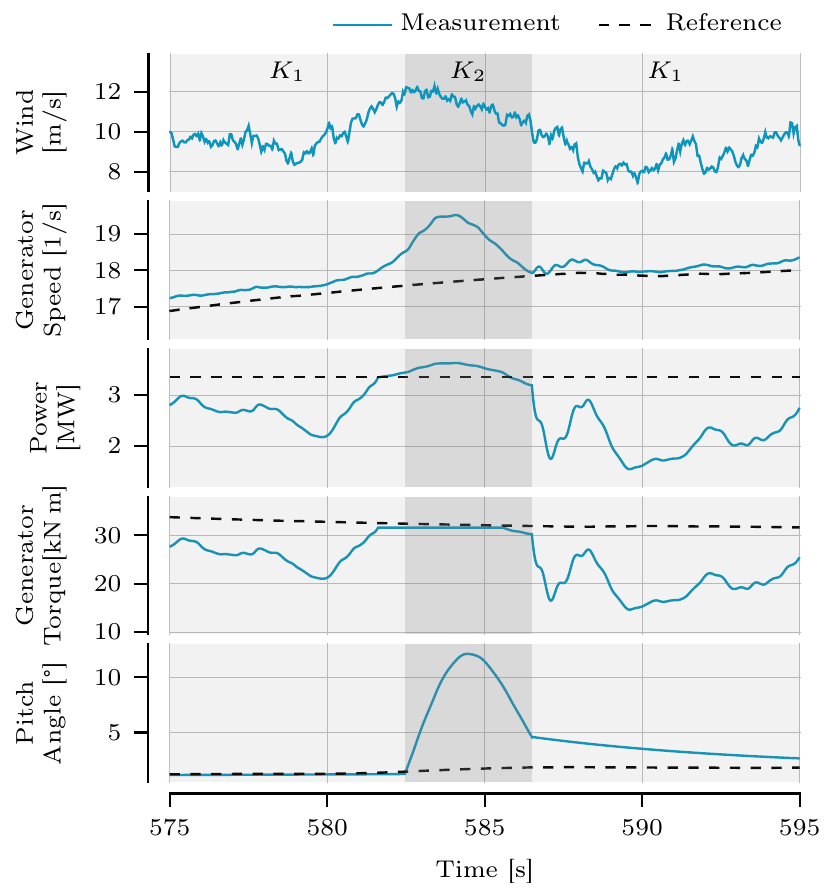}
	\caption{Measurements while switching.}
	\label{fig:switchingController}
\end{figure}
In Figure~\ref{fig:switchingController}, the wind speed, generator speed, output power, generator torque, and pitch angle are shown during the switching along with their reference trajectories.
The light gray area marks the use of $K_1$, the darker area the use of $K_2$.
At the border between both areas, the controller switches and therefore nonsmooth trajectories may occur.
Beginning at $t=\SI{579}{\second}$, the wind speed starts to increase.
Then, at $t=\SI{581}{\second}$, the generator torque reaches its maximum.
From that point on the generator speed starts to increase.
At $t=\SI{582.5}{\second}$, $\overline{V}=\SI[per-mode=fraction]{12}{\meter\per\second}$ is reached, the controller switches and immediately the pitch angle is increased to decelerate the wind turbine.
After a few seconds, the wind speed drops below $\underline{V}=\SI[per-mode=fraction]{10}{\meter\per\second}$, the controller switches back to $K_1$, and the pitch angle slowly returns to its reference value.
During the switching, no jumps occured and all trajectories were reasonably smooth.

\begin{figure}[t]
	\centering
	\includegraphics{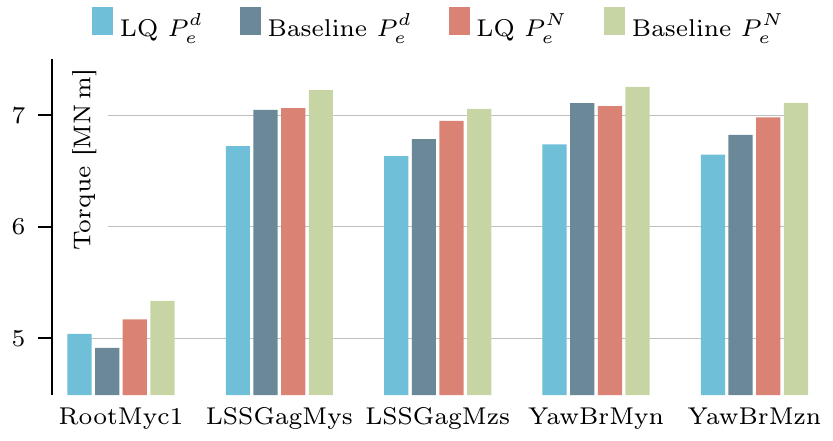}
	\caption{Damage equivalent loads for both controllers and two different power demand signals. The blade root out-of-plane bending moment (RootMyc1), the shaft non-rotating out-of-plane bending moment (LSSGagMys), the shaft non-rotating yaw bending moment (LSSGagMzs), the tower top fore-aft bending moment (YawBrMyn), and the tower top torsion bending moment (YawBrMzn).}
	\label{fig:damageEquivalentLoads}
\end{figure}

Another important aspect, when designing wind turbine controllers are mechanical loads that lead to wear.
In Figure~\ref{fig:damageEquivalentLoads}, some important \acp{del} are displayed for different scenarios.
The rainflow counting algorithm \citep{dowling1971fatigue} is employed to compute the damage amplitudes \citep{burton2011wind}.
The DELs are computed for 5 different turbine components in Figure~\ref{fig:damageEquivalentLoads}, which are prone to mechanical wear.
The baseline controller is compared with the \ac{lq} optimal controller for two different power demand scenarios.
First, a variable power demand $P^d_e$ is tracked, in the second scenario the power demand is set to the nominal $P^N = \SI{3.35}{\mega\watt}$.
In both scenarios, a wind with an average of \SI[per-mode=fraction]{15}{\meter\per\second} and a turbulence intensity of \SI{9}{\percent} is considered.
For all components except the blade root out-of-plane bending moment (RootMyc1), the LQ optimal controller achieves lower DELs for both power demand scenarios and therefore causes less mechanical wear.
The blade root out-of-plane bending moment poses an exception, where during the power tracking the LQ optimal controller leads to slightly higher DEL than the baseline controller.
However, overall an average load reduction of \SI{2.3}{\percent} could be achieved with the \ac{lq} optimal controller, while achieving comparable performance in power tracking.

% !TeX spellcheck = en_US
% !TeX encoding = UTF-8
% !TEX root = paper.tex

\section{Conclusion}
\label{sec:conclusion}
We showed that a multivariable \ac{lq} optimal control approach in combination with a relatively simple reference trajectory generation scheme achieves state-of-the-art performance with regard to power tracking.
The presented results indicate that considering the input coupling (in the \ac{lq} optimal control design) is more important than considering nonlinearities (as done by the state-of-the-art controller).
The modular control scheme allows to further improve reference trajectories which can lead to even better performance.
We also showed that the \ac{lq} optimal controller achieves lower \acp{del} throughout most simulations in comparison with the state-of-the-art.

In future work, multiple improvements are planned to be explored.
Such as relaxing the assumption that accurate wind speed measurements are available.
Moreover, the computation of the reference trajectories shall be expanded to enable better performance of the \ac{lq} optimal controller.
Finally, alternative linear multivariable control schemes will be considered.

\bibliography{literature}

\appendix
\section{Simulation Parameters}
\begin{table}[!ht]
	\centering
	\caption{IEA \SI{3.35}{\mega\watt} turbine parameters.}
	\label{tab:generalParameters}
	\begin{tabular}{c}
	\hline
    $J_t=\SI{39825631}{\kilo\gram\meter\squared}$,
    $r=\SI{65}{\meter}$,
    $\rho=\SI[per-mode=fraction]{1.225}{\kilo\gram\per\meter\cubed}$,
    $\eta=\SI{93.6}{\percent}$,\\[0.5ex]
	$N_g=\SI{97}{}$,
	$\omega_g^\text{Rated}=\SI[per-mode=fraction]{1.35}{\radian\per\second}$,
	$M_g^\text{Rated}=\SI{30.15}{\kilo\newton\meter}$,
	$\underline{\theta}=\SI{1.09}{\degree}$,\\[0.5ex]
	$\overline{\theta}=\SI{22}{\degree}$,
	$\underline{\Delta\theta}=\SI[per-mode=fraction,scientific-notation=true]{-0.000488}{\degree}$,
	$\overline{\Delta\theta}=\SI[per-mode=fraction, scientific-notation=true]{0.000488}{\degree}$,\\[0.5ex]
	$\overline{M}_g=\SI{33.17}{\kilo\newton\meter}$,
	$\underline{M}_g=\SI{0}{}$,
	$\underline{\Delta M}_g=\SI[per-mode=fraction]{-6}{\kilo\newton\meter}$,
	$\overline{\Delta M}_g=\SI[per-mode=fraction]{6}{\kilo\newton\meter}$,\\[0.5ex]
	\hline
	\end{tabular}
\end{table}

\begin{table}[!ht]
	\centering
	\caption{Baseline controller parameter configuration.}
	\label{tab:BaselineParameters}
	\begin{tabular}{c}
	\hline
	$c_{12}=\SI{82.47}{\newton\meter\second}$,
	$\omega_\text{ci}=\SI[per-mode=fraction]{10.47}{\radian\per\second}$,
	$c_M^*=\SI{1.75}{\newton\meter\second\squared}$,\\[0.5ex]
	$\omega_\text{r2}=\SI[per-mode=fraction]{15.71}{\radian\per\second}$,
	$T_M=\SI{1}{\second}$,
	$c_\theta^\text{GB}=\SI[per-mode=fraction]{30}{\radian\per\second\per\degree}$,\\[0.5ex]
	$T_\theta=\SI{0.133}{\second}$,
	$T_\text{GB}=\SI{10}{\second}$,
	$c_M^\text{GB}=\SI[per-mode=fraction]{0.0001}{\radian\per\second\per\newton\per\meter}$,\\[0.5ex]
	$K_K=\SI{0.174}{\degree}$,
	$K_I=\SI{0.004}{\degree\second}$,
	$K_P=\SI{0.133}{\degree\second}$,\\[0.5ex]
	\hline
	\end{tabular}
\end{table}

\begin{table}[!ht]
	\centering
	\caption{LQ optimal controller configuration parameters.}
	\label{tab:LQParameters}
	\begin{tabular}{c}
	\hline
	$\omega_{1}^s=\SI[per-mode=fraction]{119.31}{\radian\per\second}$,
	$\theta_1^s=\SI{2.65}{\degree}$,
	$M_{g,1}^s=\SI{16.85}{\kilo\newton\meter}$,\\[0.5ex]
	$V_1^s=\SI[per-mode=fraction]{8}{\meter\per\second}$,
	$V_2^s=\SI[per-mode=fraction]{10.5}{\meter\per\second}$,\\[0.5ex]
	$Q_1 = \operatorname{diag}(10^{-2}, 10^3, 10^3, 10^{-2})$,
	$R_1 = \operatorname{diag}(5\cdot 10^4, 5\cdot 10^4)$,\\[0.5ex]
	$\omega_{2}^s=\SI[per-mode=fraction]{119.31}{\radian\per\second}$,
	$\theta_2^s=\SI{6.98}{\degree}$,
	$M_{g,2}^s=\SI{25.72}{\kilo\newton\meter}$,\\[0.5ex]
	$Q_2 = \operatorname{diag}(10^{-4}, 10, 10^4, 10^6)$,
	$R_2 = \operatorname{diag}(10^6, 10^4)$,\\[0.5ex]
	$T_1=\SI{20}{\second}$,
	$T_2=\SI{40}{\second}$,
	$\underline{V}=\SI[per-mode=fraction]{10}{\meter\per\second}$,
	$\overline{V}=\SI[per-mode=fraction]{12}{\meter\per\second}$,\\[0.5ex]
	\hline
	\end{tabular}
\end{table}

\begin{table}[!ht]
	\centering
	\caption{Power coefficient parameters computed by fitting the function to steady-state simulation data.}
	\label{tab:coefficientParameters}
	\begin{tabular}{c}
	\hline
    $c_{1}=\SI[scientific-notation=true]{0.098}{}$,
    $c_{2}=\SI[scientific-notation=true]{-0.150}{}$,
    $c_{3}=\SI[per-mode=fraction,scientific-notation=true]{-0.011}{\per\degree}$,\\[0.5ex]
    $c_{4}=\SI[scientific-notation=true]{0.061}{}$,
    $c_{5}=\SI[per-mode=fraction,scientific-notation=true]{0.0125}{\per\degree}$,
    $c_{6}=\SI[per-mode=fraction,scientific-notation=true]{0.000053}{\per\powerTwo}$,\\[0.5ex]
    $c_{7}=\SI[scientific-notation=true]{-0.00615}{}$,
    $c_{8}=\SI[per-mode=fraction,scientific-notation=true]{-0.00184}{\per\degree}$,\\[0.5ex]
    $c_{9}=\SI[per-mode=fraction,scientific-notation=true]{-0.000338}{\per\powerTwo}$,
    $c_{10}=\SI[per-mode=fraction,scientific-notation=true]{0.0000407}{\per\powerThree}$,\\[0.5ex]
    $c_{11}=\SI[scientific-notation=true]{0.000184}{}$,
    $c_{12}=\SI[per-mode=fraction,scientific-notation=true]{0.000106}{\per\degree}$,\\[0.5ex]
    $c_{13}=\SI[per-mode=fraction,scientific-notation=true]{-0.0000515}{\per\powerTwo}$,
    $c_{14}=\SI[per-mode=fraction,scientific-notation=true]{0.0000143}{\per\powerThree}$,\\[0.5ex]
    $c_{15}=\SI[per-mode=fraction,scientific-notation=true]{-0.00000197}{\per\powerFour}$,\\[0.5ex]
	\hline
	\end{tabular}
\end{table}

\end{document}